\documentclass[%
reprint,
superscriptaddress,
 amsmath,amssymb,
 aps,
noeprints,
floats,
10pt]{revtex4-2}

\usepackage[utf8]{inputenc}
\usepackage{physics}
\usepackage{amsmath}
\usepackage{caption}
\usepackage{bm}
\usepackage{upgreek}
\usepackage{xr}
\usepackage[dvipsnames]{xcolor}
\usepackage{subcaption}
\usepackage{float}
\usepackage{tablefootnote}
\usepackage{footnote}
\usepackage{graphicx}
\usepackage{braket}
\usepackage[export]{adjustbox}
\usepackage{tabularx}
\usepackage{lipsum}  
\newcommand{\comment}[1]{}
\usepackage{multirow}
\usepackage{makecell}

\usepackage[normalem]{ulem}
\usepackage{color}
\usepackage{xcolor}

\definecolor{Mygrey}{gray}{0.80}
\definecolor{lteal}{rgb}{0.10,0.60,0.70}
\definecolor{dkred}{rgb}{0.60,0.10,0.00}
\definecolor{Navy}{rgb}{0.00,0.00,0.60}
\definecolor{Magenta}{rgb}{0.94,0.20,0.90}
\definecolor{Green}{rgb}{0.24,0.71,0.29}
\definecolor{Teal}{rgb}{0.00,0.50,0.50}

\newcommand{\bR}{{\bm R}}
\newcommand{\bP}{{\bm P}}
\newcommand{\hH}{\hat{H}}
\newcommand{{\hbG}}{{{\hat{\bm  \Gamma}}}}

\begin{document}
\pagestyle{plain}

\title{Momentum conservation for \textit{ab initio} calculations with pseudopotentials via a phase space electronic structure formulation}

\author{Guy Ohad}
    \affiliation{Department of Chemistry, Princeton University, Princeton, New Jersey 08544, United States}
\author{Nadine C.~Bradbury}
   \affiliation{Department of Chemistry, Princeton University, Princeton, New Jersey 08544, United States}
\author{Joseph E.~Subotnik}
   \affiliation{Department of Chemistry, Princeton University, Princeton, New Jersey 08544, United States}

\begin{abstract}
We demonstrate how to build a robust phase space electronic Hamiltonian $\hat{H}_\text{PS}(\bR,\bP)$ that depends on nuclear position $\bR$ and nuclear momentum $\bP$
in the presence of ionic pseudopotentials.
In particular, we show that in order to account for the non-local component of a pseudopotential while also conserving linear and angular momentum, one is forced to  introduce a non-local one-electron operator $\hbG^\text{nl}$ that couples electronic and nuclear momenta together. To test the validity of our ansatz, we derive a generalized form of Nafie's exact equality (which relates electronic momentum and electronic position) so as to include the presence of pseudopotentials. We then benchmark our proposed phase space electronic structure theory against this generalized Nafie-like result for a diverse set of diatomic molecules spanning a wide range of elements. We find excellent agreement across this {\em ab initio} test set, highlighting the strength of our method in capturing meaningful electronic momentum beyond the Born-Oppenheimer framework.
\end{abstract}

\maketitle

\section{Introduction}
\label{sec:intro}

The Born-Oppenheimer (BO) approximation is the cornerstone of modern electronic structure theory. Guided by the mass difference of nuclei versus electrons, BO theory stipulates that we freeze the nuclei when solving for the electronic stationary states of a molecular or material system; in other words, the ground state energy and molecular orbitals are functions of nuclear position $\bR$ but not nuclear momentum $\bP$. This framework is employed by every modern electronic structure package and has developed enormous predictive capability. That being said, BO theory does have several drawbacks. Perhaps most importantly, for a closed-shell molecule vibrating, a naive BO calculation always predicts that the ground-state wavefunction is real-valued, which results in  a vanishing electronic momentum.  However, this result is obviously incorrect and reflects a failure of the BO approach. For example, the electronic dipole moment  of any asymmetric diatomic (e.g.~HBr) does change in time during a vibration.

To that end, over the past few years, we have sought to go beyond the BO approach and properly account for electronic momentum by initiating so-called phase space electronic structure calculations. According to such an ansatz, we solve for the electronic stationary states of a molecule or material as a function of both $\bR$ and $\bP$. Within this framework, the electrons become aware that their underlying nuclear frame is moving such that, for $\bP \ne 0$, molecular orbitals become complex-valued (with non-zero electronic momentum). One can also show that Hamiltonian dynamics along the resulting eigensurfaces still maintain momentum conservation.  The method has shown success in recovering vibrational circular dichroism \cite{duston2024phase, tao2024electronic}, Raman optical activity \cite{tao2026non}, and Lambda-doubling \cite{peng2026lambda_doubling}. 

To date, the phase space electronic structure theory has been limited to all-electron calculations and implemented only in quantum chemistry packages with atomic-orbital basis functions. In other words, there has not yet been an implementation of phase space electronic structure theory (for any molecule or material) in an electronic structure package using a plane-wave basis. More importantly for the present work, neither the formulation nor the implementation of the theory has been extended to ionic pseudopotentials, despite their central role in the solid-state electronic structure community.

As a reminder, ionic pseudopotentials are nowadays the  standard means to treat the core electrons in \textit{ab initio} calculations of many-electron systems.  Calculations with pseudopotentials rely on the fact that core electrons usually play a marginal role in the description the physical and chemical phenomena of interest, yet they considerably increase the computational cost of the calculation. To that end, pseudopotentials replace the strong Coulomb potential of the nucleus and tightly-bound core electrons with a smoother effective potential constructed to maintain an accuracy close to that of all-electron calculations for a single atom. Consequently, pseudopotentials have become standard practice across computational condensed matter physics and quantum chemistry \cite{pickett1989pseudopotential, martin2020electronic},  offering a strong balance between accuracy and computational efficiency.

Recently, it has been noticed that fundamental complications arise when employing standard pseudopotentials for an Ehrenfest molecular dynamics simulation, stemming primarily from the non-local nature of the pseudopotential operator. Notably, rigidly displacing non-local operators alongside moving nuclei violates rototranslational sum rules and induces spurious electronic transitions. To ameliorate these problems, Stengel et al.~\cite{stengel2026rototranslational} have introduced velocity-dependent traveling pseudopotentials to restore Galilean covariance and exact sum rules in linear response and real-time dynamics. Fachin et al.~\cite{fachin2026non} have incorporated nuclear-velocity-including atomic orbitals into the projector augmented wave framework for non-adiabatic Ehrenfest dynamics with norm-conserving and ultrasoft pseudopotentials. More broadly, properly evaluating electronic momentum in \textit{ab initio} calculations with non-local pseudopotentials requires special care (see, e.g., Refs.~\cite{read1991calculation, ismail2001coupling, pickard2003nonlocal, esteve2023comprehensive}) as new velocity-dependent forces must be introduced.

Now, interestingly, while Refs.~\cite{stengel2026rototranslational, fachin2026non} address Ehrenfest theory with pseudopotentials within a complete basis and find the need for novel velocity-dependent forces, it is also known that additional forces arise even without pseudopotentials whenever we run Ehrenfest dynamics in an incomplete basis \cite{tao2024ehrenfest, amano2005quantum, krishna2007path}. 
Moreover, it is also true that running Ehrenfest dynamics in an incomplete basis can be more accurate than running Ehrenfest dynamics in a complete basis -- because Ehrenfest dynamics do not obey detailed balance and their long-time equilibrium distribution is in fact worse with more electronic states \cite{feng2026}. Thus, if we seek to restore momentum conservation in a physically meaningful manner, there are several facets that should be addressed, and the inclusion of pseudopotentials is one among several other complications.

With this context in mind, if we pivot back to the question of how to best generate potential energy surfaces,  our ansatz is that for an electronic structure framework that meaningfully incorporates electronic momentum, phase-space electronic structure theory must be appropriately reformulated in the presence of pseudopotentials (just as is done for Ehrenfest dynamics).  We hypothesize that one cannot (or at least should not) separate the problem of pseudopotentials in a moving nuclear frame from the overall larger problem of BO theory; momentum conservation should always be tackled directly in a physically consistent manner. 

Before concluding this introduction, two items must be emphasized. First, our goal below will be to develop a phase space electronic structure theory that does not rely on atomic orbitals or the introduction of explicit phases on atomically centered orbitals. As detailed in Ref.~\cite{tao2025basis}, one of the most important advances in this field has been the realization that solving the electronic structure problem in a moving frame can best be achieved by robustly partitioning space into different nuclear sectors in a basis-free fashion. This form has allowed us to perform the calculations below in a plane-wave code. 
Second, although the data below will address problems of linear momenta, our proposed Hamiltonian will conserve angular momentum, but this angular momentum takes a different form from the conserved quantity found in Ref.~\cite{stengel2026rototranslational}. Our model for angular momentum without pseudopotentials has been strongly verified by data presented in Ref.~\cite{bhati2026MSW} for a diatomic molecule.

The outline of the article below is as follows. In the remainder of this introduction, we will review phase space electronic structure theory without pseudopotentials. In Sec.~\ref{sec:theory}, we will modify the existing phase space electronic structure approach to include pseudopotentials, making sure that we maintain momentum conservation. We will find that two non-local components must be added to the existing phase space electronic structure approach, one of which corresponds exactly to the phase introduced by Stengel et al.~\cite{stengel2026rototranslational} that restored linear momentum conservation in the presence of pseudopotentials. The other contribution is novel and arises in a fashion very compatible with the existing phase space theory and is essential for angular momentum conservation. Moreover, we revisit Nafie's equality, relating electronic momentum to electronic position, as means to validate our phase space approach. Thereafter, we present the methods for our calculations in Sec.~\ref{sec:methods}. The results are then shown in Sec.~\ref{sec:results} where we demonstrate that, for a large set of diatomic molecules, we do in fact reproduce Nafie's equality very well. We conclude in Sec.~\ref{sec:conclusions} and highlight future directions.

\subsection{Brief overview of phase space electronic structure theory for all-electron problems}

Before presenting the salient equations, we establish our notation. Nuclei are indexed by $I$ and $J$, with $\bm R_I$, $\bm P_I$, and $M_I$ denoting the position, momentum, and mass of nucleus $I$, respectively. Electronic states are indexed by $i,j$; $\hat{\bm r}$, $\hat{\bm p}$, $\hat{\bm l}$, and $\hat{\bm s}$ denote the \emph{total} electronic position, linear momentum, orbital angular momentum, and spin, respectively, and $m_e$ denotes the electron mass. Operators are indicated by a hat, and three-dimensional vectors by boldface. Cartesian components are indexed $\alpha$, $\beta$, and $\gamma$.

The central goal of phase space electronic structure approaches \cite{bian2026phase} is to account for non-adiabatic electron-nuclear coupling effects beyond the BO approximation. As inspired by Shenvi's phase space surface hopping paper \cite{shenvi2009phase}, when working with the exact many-body Hamiltonian for electrons and nuclei in the adiabatic electronic basis, one treats nuclear momentum and position as classical parameters rather than quantum operators (formally, we now understand that such ``classicalization'' is exactly a Wigner transformation over the nuclei, though that was not the language adopted by Shenvi). After transforming back to the diabatic representation, the Wignerized Hamiltonian reads
\begin{align}
    \label{eq:H_Shenvi}
    \hat{H}_W = \sum_I \frac{1}{2M_I} \left( \bm P_I -i \hbar \hat{\bm d}^I (\bm R) \right)^2 + \hat{H}_\text{el}\left( \bm R \right),
\end{align}
where the derivative coupling vector operator associated with nucleus $I$, $\hat{\bm d}^I$, satisfies
\begin{align}
    \label{eq:derivative_coupling}
    \Braket{{\Phi_i}|\hat{\bm d}^I|{\Phi_j}} = \Braket{\Phi_i | \frac{\partial \Phi_j}{\partial \bm{R}_I}} \equiv \bm{d}_{ij}^I.
\end{align}
The BO states $\Ket{\Phi_i \left( \bm R \right)}$ are defined as eigenstates of the BO (or electronic) Hamiltonian $\hat{H}_\text{el}$ with eigenvalues $E_i\left( \bm R \right)$ for a given nuclear configuration, with  
\begin{align}
    \label{eq:H_el}
    \hat{H}_\text{el}\left( \bm R \right) = \frac{\hat{\bm p}^2}{2m_e} + \hat{V}_\text{en} + \hat{V}_\text{nn} + \hat{V}_\text{ee}. 
\end{align}
Here the potential terms represent electron-nuclear (en), nuclear-nuclear (nn), and electron-electron (ee) interactions.

The derivative coupling vectors encode all the beyond BO information on coupling between electronic states and nuclear motion. However, while Eq.~\eqref{eq:H_Shenvi} is the exact Wigner transform of the Hamiltonian, running classical nuclear dynamics along eigenvalues of $\hH_W$ can be very problematic; the dynamics are numerically unstable near electronic curve crossings, they are prohibitively expensive computationally for most real systems, and they are ill defined for problems with degeneracy (e.g.~spin degrees of freedom). These obstacles have motivated the more recent development of new phase space electronic structure approaches that approximate the derivative coupling vector as a one-electron operator $\hat{\bm \Gamma}_I$ \cite{bian2026phase, subotnik2026perspective}. 

While there can be many ways to choose the form of $\hat{\bm \Gamma}_I$, it would be appealing to construct an operator that satisfies the same symmetry constraints as does the exact derivative coupling vectors. We denote this choice of $\hat{\bm \Gamma}_I$ by $\hat{\bm \Gamma}_I^\text{local}$, a terminology whose motivation will become apparent below. $\hat{\bm \Gamma}_I^\text{local}$ is built to satisfy four exact conditions that reflect translational and rotational invariance of the entire system as well as the correct transformation of $\hat{\bm \Gamma}_I^\text{local}$ under translation and rotation. These conditions ensure conservation of the total linear and angular momentum of the system and are as follows:
\begin{align}
    \label{eq:Gamma_constr1}
    -i \hbar \sum_I \hat{\bm \Gamma}_I^\text{local} + \hat{\bm p} = 0, 
\end{align}
\begin{align}
    \label{eq:Gamma_constr2}
    \left[ -i \hbar \sum_I \frac{\partial}{\partial \bm R_I} + \hat{\bm p}, \hat{\bm \Gamma}_J^\text{local}  \right] = 0,
\end{align}
\begin{align}
    \label{eq:Gamma_constr3}
    -i \hbar \sum_I \left( \bm R_I \times \hat{\bm \Gamma}_I^\text{local} \right) + \hat{\bm l} + \hat{\bm s} = 0, 
\end{align}
\begin{align}
    \label{eq:Gamma_constr4}
    \left[ -i \hbar \sum_I \left( \bm R_I \times \frac{\partial}{\partial \bm R_I} \right)_\alpha + \hat{l}_\alpha + \hat{s}_\alpha, \hat{\Gamma}_{J \beta}^\text{local}  \right] \nonumber \\
    = i \hbar \sum_\gamma \varepsilon_{\alpha \beta \gamma}  \hat{\Gamma}_{J \gamma}^\text{local}.
\end{align}
The origin and proof of these conditions are discussed elsewhere \cite{wu2024linear, littlejohn2023representation, tao2025basis}.

Following recent developments \cite{tao2025basis,bian2026phase, subotnik2026perspective}, $\hat{\bm \Gamma}_I^\text{local}$ is constructed as a sum of two components
\begin{align}
    \label{eq:Gamma_local}
    \hat{\bm \Gamma}_I^\text{local} = \hat{\bm \Gamma}^{\prime \text{local}}_I + \hat{\bm \Gamma}^{\prime \prime \text{local}}_I.
\end{align}
The first component, $\hat{\bm \Gamma}^{\prime \text{local}}_I$, is based on electronic translation factors \cite{tao2025basis} but crucially does not require any definition of an atomic orbital basis set. We set:
\begin{align}
    \label{eq:Gamma_prime_local}
    \hat{\bm \Gamma}^{\prime \text{local}}_I = \frac{1}{2i \hbar} \left( \hat{\theta}_I \hat{\bm p} + \hat{\bm p} \hat{\theta}_I \right)
\end{align}
where $\hat{\theta}_I$ is any partition of unity of space that satisfies
\begin{align}
    \sum_I \theta_I \left( \bm r \right) = 1.
\end{align}
A proper choice of $\hat{\theta}_I$ is such that it transitions smoothly between zero far from the position of nucleus $I$ and unity near it, such that $\hat{\bm \Gamma}^{\prime \text{local}}_I$ can intuitively be understood as the local electronic momentum around nucleus $I$. There have been debates on the best approach for choosing $\hat{\theta}_I$ \cite{bian2025phase, tao2025basis}, and the one employed in this work is detailed in the Methods section.

The second component, $\hat{\bm \Gamma}^{\prime \prime \text{local}}_I$, is based on electronic rotation factors \cite{qiu2024simple} and reads:
\begin{align}
    \label{eq:Gamma_double_prime_local}
    \hat{\bm \Gamma}^{\prime \prime \text{local}}_I = \sum_J \zeta_{IJ} \left( \bm R_I - \bm R_J^0 \right) \times \left( \bm K_J^{-1} \hat{\bm{\mathcal{J}}}_J \right),
\end{align}
where
\begin{align}
    \hat{\bm{\mathcal{J}}}_J \equiv \frac{1}{2i\hbar} \left( \left( \hat{\bm r} - \bm R_J \right) \times \hat{\theta}_J \hat{\bm p} + \left( \hat{\bm r} - \bm R_J \right) \times \hat{\bm p} \hat{\theta}_J \right).
\end{align}
Here,
\begin{align}
    \bm R_J^0 = \frac{\sum_I \zeta_{IJ} \bm R_I}{\sum_I \zeta_{IJ}}
\end{align}
is a local average position and
\begin{align}
    \bm K_J =  \sum_I \zeta_{IJ} \left( \bm R_I \bm R_I^T - \bm R_J^0 \bm R_J^{0T} - \left( \bm R_I^T \bm R_I - \bm R_J^{0T} \bm R_J^0  \right) \mathcal{I}_3  \right)
\end{align}
is a local moment of inertia. The localization function
\begin{align}
    \zeta_{IJ} = M_I \text{exp} \left( - \frac{\abs{\bm R_I - \bm R_J}^2}{\sigma_{IJ}} \right)
\end{align}
controls the local coupling between nuclear rotations and electronic angular momentum via a length scale parameter $\sigma_{IJ}$. This length scale reflects the degree of electronic communication between atoms $I$ and $J$, and whether or not an electron in the vicinity of atom $I$ experiences a Coriolis force from the relative motion of atom $J$. $\mathcal{I}_3$ is the $3 \times 3$ identity matrix.

One can show that $\hat{\bm \Gamma}_I^\text{local}$ in Eq.~\eqref{eq:Gamma_local} satisfies conditions \eqref{eq:Gamma_constr1}-\eqref{eq:Gamma_constr4}. When put in place of $\hat{\bm d}^I$ in Eq.~\eqref{eq:H_Shenvi}, the Hamiltonian can be diagonalized at a cost similar to diagonalizing the BO Hamiltonian. The framework just described has now been studied in many different applications, with results consistently outperforming BO theory \cite{duston2024phase, tao2024electronic, bian2025phase, tao2026non, zaidi2026electron, peng2026lambda_doubling, bhati2026MSW, peng2026conceptual, zaidi2026exact}.

\section{Theory: The Electronic Phase Space Hamiltonian With Pseudopotentials}
\label{sec:theory}

Our motivation here is to revisit the phase space electronic structure approach above in the presence of pseudopotentials. From here onward, we consider the case where the true many-electron and many-nuclei electron-nuclear Coulomb potential $\hat{V}_\text{en}$ in Eq.~\eqref{eq:H_el} is approximated with pseudopotentials, i.e.~it is replaced with
\begin{align}
    \hat{V}^\text{pseudo} = \sum_I \left( \hat{v}_I^\text{local} + \hat{v}_I^\text{nl} \right),
\end{align}
where $\hat{v}_I^\text{local}$ is a fully local (multiplicative) potential in real space and $\hat{v}_I^\text{nl}$ is a non-local (non-multiplicative) potential. Taken together, they constitute the pseudopotential for atom $I$.

\subsection{Canonical versus kinetic momentum}

In the presence of non-local pseudopotentials, the canonical momentum $\hat{\bm p}$ is no longer the momentum naturally identified through the usual Schr\"{o}dinger equation of motion. Since the non-local potential does not commute with position operator, the electronic velocity is not simply $\hat{\bm p}/m_e$. The corresponding physical momentum is instead the electronic \emph{kinetic} momentum, $\bm{\hat{\pi}}$, defined as 
\begin{align}
    \hat{\bm \pi} &\equiv i\frac{m_e}{\hbar} \left[ \hat{H}_\text{el}, \hat{\bm r} \right] \label{eq:pi_definition} \\  &= \hat{\bm p} + i\frac{m_e}{\hbar} \left[ \sum_I \hat{v}_I^\text{nl}, \hat{\bm r} \right]. \label{eq:pi_in_terms_of_p}
\end{align}
Similarly, we define the electronic kinetic angular momentum
\begin{align}
    \hat{\bm \lambda} &\equiv \hat{\bm r} \times \hat{\bm \pi} \label{eq:lambda_def} \\
    &= \hat{\bm l} + i\frac{m_e}{\hbar} \hat{\bm r} \times \left[ \sum_I \hat{v}_I^\text{nl}, \hat{\bm r} \right],
\end{align}
where $\hat{\bm l} = \hat{\bm r} \times \hat{\bm p}$. A natural extension of the phase space electronic structure approach described above (without pseudopotentials) is then to conserves the total kinetic linear momentum
\begin{align}
    \label{eq:total_kinetic_linear_momentum}
    \bm \Pi_\text{tot} \equiv \sum_I M_I \dot{\bm R}_I + \Braket{\hat{\bm \pi}}
\end{align}
and total kinetic angular momentum
\begin{align}
    \label{eq:total_kinetic_angular_momentum}
    \bm \Lambda_\text{tot} \equiv \sum_I M_I \bm R_I \times \dot{\bm R}_I + \Braket{\hat{\bm \lambda} + \hat{\bm s}}.
\end{align}

We note that the quantity in Eq.~\eqref{eq:total_kinetic_linear_momentum} corresponds exactly to the conserved linear momentum with traveling pseudopotentials as introduced by Stengel et al.~\cite{stengel2026rototranslational}. The quantity in Eq.~\eqref{eq:total_kinetic_angular_momentum}, on the other hand, is different from the corresponding angular momentum associated with traveling pseudopotentials; it represents a fundamental physical quantity. Interestingly, $\hat{\bm \lambda}$  in Eq.~\eqref{eq:lambda_def} is consistent with the Ismail-Beigi et al.~\cite{ismail2001coupling} first order correction for magnetic observables which restores gauge invariance in the presence of general external fields and non-local potentials. In such a formulation, $\hat{\bm \lambda}$ arises from a derivative of the Hamiltonian with respect to a uniform magnetic field. 

We emphasize that, in the all-electron phase space picture presented in the introduction, the electronic canonical and kinetic momenta are one and the same, and $\hat {\bm \Gamma}_I^\text{local}$ is sufficient for momentum conservation. In such a case, the total linear momentum is $\bm P_\text{tot} = \sum_I M_I \dot{\bm R}_I + \Braket{\hat{\bm p}}$ and the total angular momentum is $\bm L_\text{tot} = \sum_I M_I \bm R_I \times \dot{\bm R}_I + \Braket{\hat{\bm l} + \hat{\bm s}}$. 
This scenario is no longer the case when pseudopotentials enter the calculation.

\subsection{A non-local addition to the $\hat{\bm \Gamma}$ operator}
\label{subsec:adding_Gamma_nl}

In order to conserve $\bm \Pi_\text{tot}$ and $ \bm \Lambda_\text{tot}$, we hypothesize that the proper phase space electronic structure Hamiltonian takes the form:
\begin{align}
    \label{eq:HPS_Gamma_nl}
    \hat{H}_\text{PS} = \sum_I \frac{1}{2M_I} \left( \bm P_I -i \hbar \hat{\bm \Gamma}_I \right)^2 + \hat{H}_\text{el}\left( \bm R \right),
\end{align}
where
\begin{align}
    \label{eq:new_gamma_with_nl_component} 
    \hat{\bm \Gamma}_I &= \hat{\bm \Gamma}_I^\text{local} +\hat{\bm \Gamma}_I^\text{nl}.   
\end{align}
Furthermore, we submit that it is natural to split $\hat{\bm \Gamma}_I^\text{nl}$ in the same fashion as $\hat{\bm \Gamma}_I^\text{local}$, namely
\begin{align}
    \label{eq:Gamma_nl}
    \hat{\bm \Gamma}_I^\text{nl} = \hat{\bm \Gamma}^{\prime \text{nl}}_I + \hat{\bm \Gamma}^{\prime \prime \text{nl}}_I,
\end{align}
where
\begin{align}
    \hat{\bm \Gamma}^{\prime \text{nl}}_I & \equiv  \frac{m_e}{\hbar^2} \left[ \hat{v}_I^\text{nl}, \hat{\bm r} \right]
\end{align}
and
\begin{align}
    \hat{\bm \Gamma}^{\prime \prime \text{nl}}_I \equiv \sum_J \zeta_{IJ} \left( \bm R_I - \bm R_J^0 \right) \times \left( \bm K_J^{-1} \hat{\bm{\mathcal{J}}}_J^\text{nl} \right),
\end{align}
\begin{align}
    \label{eq:J_nl}
    \hat{\bm{\mathcal{J}}}_J^\text{nl} = \frac{m_e}{\hbar^2} \left( \hat{\bm r} - \bm R_J \right) \times \left[ \hat{v}_J^\text{nl}, \hat{\bm r} \right].
\end{align}
Eqs.~\eqref{eq:new_gamma_with_nl_component}-\eqref{eq:J_nl} constitute the central results of this work. As shown below, this form of $\hat{\bm \Gamma}_I$ guarantees momentum conservation. Before proving so, we note that $\hat{\bm \Gamma}_I$ obeys the conditions
\begin{align}
    \label{eq:Gamma_constr1_new}
    -i \hbar \sum_I \hat{\bm \Gamma}_I + \hat{\bm \pi} = 0, 
\end{align}
\begin{align}
    \label{eq:Gamma_constr2_new}
    \left[ -i \hbar \sum_I \frac{\partial}{\partial \bm R_I} + \hat{\bm p}, \hat{\bm \Gamma}_J  \right] = 0,
\end{align}
\begin{align}
    \label{eq:Gamma_constr3_new}
    -i \hbar \sum_I \left( \bm R_I \times \hat{\bm \Gamma}_I \right) + \hat{\bm \lambda} + \hat{\bm s} = 0, 
\end{align}
\begin{align}
    \label{eq:Gamma_constr4_new}
    \left[ -i \hbar \sum_I \left( \bm R_I \times \frac{\partial}{\partial \bm R_I} \right)_\alpha + \hat{l}_\alpha + \hat{s}_\alpha, \hat{\Gamma}_{J \beta}  \right] \nonumber \\
    = i \hbar \sum_\gamma \varepsilon_{\alpha \beta \gamma}  \hat{\Gamma}_{J \gamma}.
\end{align}
Eqs.~\eqref{eq:Gamma_constr1_new} and \eqref{eq:Gamma_constr3_new} follow trivially from the corresponding conditions for $\hat{\bm \Gamma}_I^\text{local}$ and the definition of $\hat{\bm \pi}$ and $\hat{\bm \lambda}$. Eqs.~\eqref{eq:Gamma_constr2_new} and \eqref{eq:Gamma_constr4_new} are proven in Appendix A. Note that the derivative coupling operator $\hat{\bm d}^I$ does not satisfy Eqs.~\eqref{eq:Gamma_constr1_new}-\eqref{eq:Gamma_constr4_new}; instead, it satisfies Eqs.~\eqref{eq:Gamma_constr1}-\eqref{eq:Gamma_constr4}. 

\subsection{Momentum conservation}
\label{subsec:momentum_conservation}

To prove momentum conservation, we start from Hamilton's equations of motion for classical nuclei along a given phase space surface
\begin{align}
    \label{eq:hamilton_R}
    \dot{\bm R}_I = \frac{\partial E_i^\text{PS}}{\partial \bm P_I} = \Braket{\Psi_i^\text{PS} | \frac{\partial \hat{H}_\text{PS}}{\partial \bm P_I} | \Psi_i^\text{PS}},
\end{align}
\begin{align}
    \label{eq:hamilton_P}
    \dot{\bm P}_I = - \frac{\partial E_i^\text{PS}}{\partial \bm R_I} = - \Braket{\Psi_i^\text{PS} | \frac{\partial \hat{H}_\text{PS}}{\partial \bm R_I} | \Psi_i^\text{PS}}
\end{align}
where $E_i^\text{PS}$ and $\Ket{\Psi_i^\text{PS}}$ are the eigenenergies and eigenstates, respectively, of $\hat{H}_\text{PS}$. The second equality in \eqref{eq:hamilton_R} and \eqref{eq:hamilton_P} follows from the Hellmann-Feynman theorem.

Employing Eq.~\eqref{eq:hamilton_R} in Eqs.~\eqref{eq:total_kinetic_linear_momentum}-\eqref{eq:total_kinetic_angular_momentum}, we obtain
\begin{align}
    \label{eq:pi_tot_reduced}
    \bm \Pi_\text{tot} &= \sum_I \bm P_I - i \hbar  \sum_I \Braket{{\Psi_i^\text{PS}}|\hat{\bm \Gamma}_I|\Psi_i^\text{PS}} +\Braket{{\Psi_i^\text{PS}}|\hat{\bm \pi}|\Psi_i^\text{PS}} \nonumber \\
    & = \sum_I \bm P_I,
\end{align}
\begin{align}
\label{eq:lambda_tot_reduced}
\bm{\Lambda}_\text{tot}
&=
\begin{aligned}[t]
&\sum_I \bm R_I \times \bm P_I
-i\hbar \sum_I \bm R_I \times
\Braket{\Psi_i^\text{PS}|\hat{\bm \Gamma}_I|\Psi_i^\text{PS}} \nonumber \\
&+ \Braket{\Psi_i^\text{PS}|
\hat{\bm \lambda} + \hat{\bm s}
|\Psi_i^\text{PS}}
\end{aligned}
\\
&= \sum_I \bm R_I \times \bm P_I,
\end{align}
where we used \eqref{eq:Gamma_constr1_new} and \eqref{eq:Gamma_constr3_new}. Taking the time derivative of \eqref{eq:pi_tot_reduced} and \eqref{eq:lambda_tot_reduced}, we find
\begin{align}
    \label{eq:Pi_tot_mid}
    \dot{\bm \Pi}_\text{tot}=  - \Braket{\Psi_i^\text{PS} | \sum_I \frac{\partial \hat{H}_\text{PS}}{\partial \bm R_I} | \Psi_i^\text{PS}},
\end{align}
\begin{align}
    \label{eq:lambda_tot_mid}
    \dot{\bm \Lambda}_\text{tot}= &- \sum_I \bm P_I \times \Braket{\Psi_i^\text{PS} |  \frac{\partial \hat{H}_\text{PS}}{\partial \bm P_I} | \Psi_i^\text{PS}} \nonumber \\
    &- \sum_I \bm R_I \times \Braket{\Psi_i^\text{PS} |  \frac{\partial \hat{H}_\text{PS}}{\partial \bm R_I} | \Psi_i^\text{PS}}.
\end{align}
Now, the electronic Hamiltonian is translationally (rotationally) invariant upon translating (rotating) all nuclei and all electrons together (e.g., by an infinitesimal amount). Mathematically, this invariance is expressed by
\begin{align}
    \label{eq:trans_inv_Hel}
    \left[ -i \hbar \sum_I \frac{\partial}{\partial \bm R_I} + \hat{\bm p}, \hat{H}_\text{el}  \right] = 0,
\end{align}
\begin{align}
    \label{eq:rotat_inv_Hel}
    \left[ -i \hbar \sum_I \left(\bm R_I \times \frac{\partial}{\partial \bm R_I} +\bm P_I \times \frac{\partial}{\partial \bm P_I} \right) + \hat{\bm l}  + \hat{\bm s}, \hat{H}_\text{el}  \right] = 0.
\end{align}
Thus, provided that $\hat{\bm \Gamma}_I$ satisfies conditions \eqref{eq:Gamma_constr2_new} and \eqref{eq:Gamma_constr4_new}, it follows that $\hat{H}_\text{PS}$ is also translationally and rotationally invariant and satisfies commutation relations analogous to Eqs.~\eqref{eq:trans_inv_Hel} and \eqref{eq:rotat_inv_Hel}. These commutation relations can also be written as
\begin{align}
    \label{eq:trans_inv_HPS}
    \sum_I \frac{\partial \hat{H}_\text{PS}}{\partial \bm R_I} = -\frac{i}{\hbar} \left[ \hat{\bm p} , \hat{H}_\text{PS} \right],
\end{align}
\begin{align}
    \label{eq:rotat_inv_HPS}
    \sum_I \left(\bm R_I \times \frac{\partial \hat{H}_\text{PS}}{\partial \bm R_I} +\bm P_I \times \frac{\partial \hat{H}_\text{PS}}{\partial \bm P_I} \right) = -\frac{i}{\hbar} \left[ \hat{\bm l} + \hat{\bm s}, \hat{H}_\text{PS} \right].
\end{align}
Finally, momentum conservation follows from plugging Eq.~\eqref{eq:trans_inv_HPS} in Eq.~\eqref{eq:Pi_tot_mid}, as well as plugging Eq.~\eqref{eq:rotat_inv_HPS} in Eq.~\eqref{eq:lambda_tot_mid}. Recalling that $\Ket{\Psi_i^\text{PS}}$ is an eigenstate of $\hat{H}_\text{PS}$, we find
\begin{align}
    \dot{\bm \Pi}_\text{tot} &=  \frac{i}{\hbar} \Braket{\Psi_i^\text{PS} | \left[ \hat{\bm p} , \hat{H}_\text{PS} \right] | \Psi_i^\text{PS}} \nonumber \\
    &= 0,
\end{align}
\begin{align}
    \dot{\bm \Lambda}_\text{tot} &=  \frac{i}{\hbar} \Braket{\Psi_i^\text{PS} | \left[ \hat{\bm l} +\hat{\bm s}, \hat{H}_\text{PS} \right] | \Psi_i^\text{PS}} \nonumber \\
    &= 0,
\end{align}
yielding the desired conservation laws.
Importantly, note that the cancellation in Eqs.~\eqref{eq:pi_tot_reduced} and \eqref{eq:lambda_tot_reduced} above occurs only because we include $\hat{\bm \Gamma}_I^\text{nl}$ in the phase space Hamiltonian; momentum conservation requires that the form of $\hat{\bm \Gamma}_I$ account for the non-local pseudopotential.

\begin{figure*}[htbp!]
    \centering
    \includegraphics[width=\linewidth]{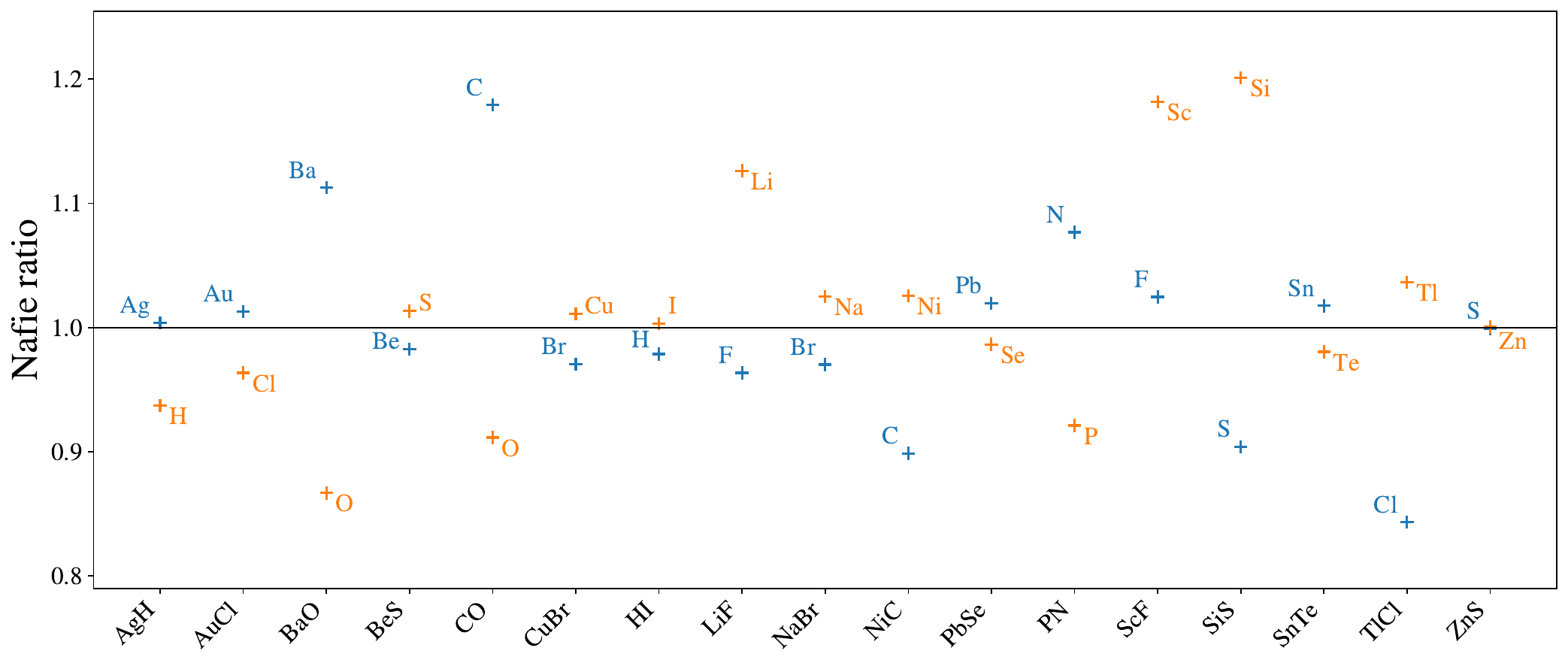}
    \caption{The Nafie ratio, $\Braket{\hat{\bm \pi}}_{\Psi_0^\text{PS}} \Big/ \left( m_e \frac{d\Braket{\hat{\bm r}}_{\Phi_0}}{dt} + m_e \Braket{\hat{\bm \eta}^\text{nl}}_{\Phi_0} \right)$, computed for a set of 17 diatomic molecules. Two vibrational modes are calculated for each molecule, represented by one blue and one orange cross above each entry in the horizontal axis. The label near the cross indicates the translated atom, while the other atom remains stationary. The solid black line corresponds to the ideal ratio of 1.}
    \label{fig:nafie_ratio}
\end{figure*}

Lastly, note that if we did not include $\hat{\bm \Gamma}^{\prime \prime \text{nl}}_I$, we would instead conserve the quantity
\begin{align}
    \Tilde{\bm L}_\text{tot} = \bm L_\text{tot} + i \frac{m_e}{\hbar}\sum_I \bm R_I \times \Braket{\left[ \hat{v}_I^\text{nl}, \hat{\bm r} \right]},
\end{align}
which aligns with the conserved angular momentum with traveling pseudopotentials (see Eqs.~(9)-(10) in Ref.~\cite{stengel2026rototranslational}).

\subsection{Nafie's equality}

As mentioned above, a significant improvement of the phase space electronic structure approach over BO is the recovery of a meaningful, non-zero electronic momentum. One means of checking the validity of phase space electronic structure theory is to compare its predictions against Nafie's equality \cite{nafie1983adiabatic, tao2024practical}. As a reminder, the latter is obtained from perturbing the electronic Hamiltonian with an electron-nuclear coupling term that incorporates the exact derivative coupling operator $\hat{\bm d}^I$. In the all-electron picture, Nafie's relation is a very intuitive, Ehrenfest-theorem-like expression
\begin{align}
    \label{eq:nafie_orig}
    \Braket{\Psi_i | \hat{\bm p} | \Psi_i} = m_e \frac{d \Braket{\Phi_i| \hat{\bm r}| \Phi_i}}{dt},
\end{align}
where $\Ket{\Psi_i}$ is an eigenstate of the perturbed Hamiltonian.  Ref.~\cite{tao2025basis} has demonstrated the success of the phase space approach in recovering electronic momentum with the correct sign as benchmarked against Nafie's equality in Eq.~\eqref{eq:nafie_orig}, 
an empirical fact 
that validates the choice of  $\hat{\bm \Gamma}_I$ as a meaningful approximation to $\hat{\bm d}^I$ in the absence of pseudopotentials.

Before running and testing any phase space electronic structure calculations against this relation, let us revisit the derivation of Nafie's equality for the case of pseudopotentials. In accordance with our suggested phase space electronic structure picture, the starting point is the Hamiltonian
\begin{align}
    \label{eq:H_Nafie}
    \hat{H}_\text{Nafie}=\hat{H}_\text{el} -i \hbar \sum_I \frac{\bm P_I \cdot \left( \hat{\bm d}^I +\hat{\bm \Gamma}_I^\text{nl} \right)}{M_I}.
\end{align}
The above Hamiltonian includes the electron-nuclear coupling term from the Hamiltonian in Eq.~\eqref{eq:H_Shenvi}, with the additional ingredient $\hat{\bm \Gamma}_I^\text{nl}$ so as to account for the presence of pseudopotentials in $\hat{H}_\text{el}$. As stated in Sec.~\ref{subsec:adding_Gamma_nl}, even in the presence of pseudopotentials, the true derivative coupling $\hat{\bm d}^I$ in Eq.~\eqref{eq:derivative_coupling} still satisfies conditions \eqref{eq:Gamma_constr1}-\eqref{eq:Gamma_constr4} (replacing $\hat{\bm \Gamma}_I^\text{local}$ with $\hat{\bm d}^I$). However, as stated in Sec.~\ref{subsec:momentum_conservation}, for any dynamics involving coupled nuclei and electrons to obey momentum conservation, the relevant Hamiltonian in Eq.~\eqref{eq:H_Nafie} must include the factor $\hat{\bm \Gamma}_I^\text{nl}$. Thus, in the presence of pseudopotentials, the form of the non-BO couplings has necessarily changed and can no longer be the simple factor of $\frac{\bm P \cdot \hat{\bm d}}{M}$ that we normally expect.

Following Nafie, we treat the second term on the right hand side of Eq.~\eqref{eq:H_Nafie} as a perturbation to $\hat{H}_\text{el}$. The perturbed wavefunctions, to first order, are then given by
\begin{align}
    \Ket{\Psi_i} = \Ket{\Phi_i} - i \hbar \sum_{j \ne i} \frac{\Braket{\Phi_j | \sum_I \frac{\bm P_I \cdot \left( \hat{\bm d}^I +\hat{\bm \Gamma}_I^\text{nl} \right)}{M_I} | \Phi_i}}{E_i-E_j}.
\end{align}
Let $\left( \bm \Gamma_I^\text{nl} \right)_{ji} \equiv \Braket{\Phi_j|\hat{\bm \Gamma}_I^\text{nl}|\Phi_i}$. One can then evaluate
\begin{align}
   & \Braket{\Psi_i | \hat{\bm \pi} | \Psi_i}\\
   &= 2m_e \text{Re} \sum_{j\ne i,I} \frac{\bm P_I}{M_I} \cdot
   \frac{\bm d^I_{ji} + \left( \bm \Gamma_I^\text{nl} \right)_{ji}}{E_i-E_j}
   \Braket{\Phi_i | \left[ \hat{H}_\text{el}, \hat{\bm r} \right] | \Phi_j} \\
   &= 2m_e\text{Re} \sum_{jI} \Braket{\Phi_i | \hat{\bm r} | \Phi_j}
   \Braket{\Phi_j | \frac{\bm P_I}{M_I} \cdot
   \left( \frac{\partial}{\partial \bm R_I} + \hat{\bm \Gamma}_I^\text{nl} \right)
   | \Phi_i} \\
   &= 2m_e \text{Re}
   \Braket{\Phi_i | \hat{\bm r} \sum_I \frac{d \bm R_I}{dt}
   \cdot \frac{\partial}{\partial \bm R_I} | \Phi_i} \\
   &\phantom{={}} {}+ 2m_e \text{Re}
   \Braket{\Phi_i | \hat{\bm r} \sum_I \frac{\bm P_I}{M_I}
   \cdot \hat{\bm \Gamma}_I^\text{nl} | \Phi_i} \\
   &= 2m_e \text{Re}
   \Braket{\Phi_i | \hat{\bm r} \frac{d}{dt}| \Phi_i} \\
   &\phantom{={}} {}+ 2m_e \text{Re}
   \Braket{\Phi_i | \hat{\bm r} \sum_I \frac{\bm P_I}{M_I}
   \cdot \hat{\bm \Gamma}_I^\text{nl} | \Phi_i}
\end{align}
\begin{align}
    \Braket{\Psi_i | \hat{\bm \pi} | \Psi_i}= &m_e \frac{d \Braket{\Phi_i | \hat{\bm r}| \Phi_i}}{dt} \nonumber \\
    &+m_e \Braket{\Phi_i | \left[ \hat{\bm r} , \sum_I \frac{\bm P_I}{M_I} \cdot \hat{\bm \Gamma}_I^\text{nl} \right] | \Phi_i}. \label{eq:nafie_relation_extended}
\end{align}
Here we have used $\bm d^I_{ii}=0$ and $\left( \bm \Gamma_I^\text{nl} \right)_{ii} = 0 $ if the BO states are chosen real \footnote{The expectation value of an anti-Hermitian and real operator for a real state vanishes.} and $\dot{\bm R}_I = \frac{\bm P_I}{M_I}$ for dynamics along a BO surface. We submit that Eq.~\eqref{eq:nafie_relation_extended} is the analogue of Nafie's equality in the presence of pseudopotentials.

Finally, note that in the absence of non-local pseudopotentials, $\hat{\bm \pi}$ and $\hat{\bm p}$ are one and the same and the second term on the right hand side of Eq.~\eqref{eq:nafie_relation_extended} vanishes. In this case, Eq.~\eqref{eq:nafie_relation_extended} reduces to Nafie's original result in Eq.~\eqref{eq:nafie_orig}.

\section{Methods: Testing Nafie's equality}
\label{sec:methods}

To test the robustness of our theory, we examine the validity of Nafie's equality, Eq.~\eqref{eq:nafie_relation_extended}, in \textit{ab-initio} calculations that employ pseudopotentials. We consider a set 17 diatomic molecules for our calculations: AgH, AuCl, BaO, BeS, CO, CuBr, HI, LiF, NaBr, NiC, PbSe, PN, ScF, SiS, SnTe, TlCl, and ZnS. This set was chosen to span a broad range of elements and pseudopotentials with different core–valence partitioning. In our calculations, we test whether
\begin{align}
    \label{eq:nafie_ab_initio}
     \Braket{\hat{\bm \pi}}_{\Psi_0^\text{PS}} \overset{?}{=} m_e \frac{d\Braket{\hat{\bm r}}_{\Phi_0}}{dt} + m_e \Braket{\hat{\bm \eta}^\text{nl}}_{\Phi_0},
\end{align}
where the subscript $0$ represents the ground state. More specifically, our diagnostic is the Nafie ratio, which we define as the ratio between the left- and right-hand sides of Eq.~\eqref{eq:nafie_ab_initio}, namely
\begin{align}
\label{eq:nafie_ratio_definition}
    \text{Nafie ratio} \equiv \frac{\Braket{\hat{\bm \pi}}_{\Psi_0^\text{PS}}}{m_e \frac{d\Braket{\hat{\bm r}}_{\Phi_0}}{dt} + m_e \Braket{\hat{\bm \eta}^\text{nl}}_{\Phi_0}} \overset{?}{=} 1
\end{align}
Here, we have used the following short hand notations:
\begin{align}
   \Braket{\hat{\bm \pi}}_{\Psi_0^\text{PS}} &\equiv \Braket{\Psi_0^\text{PS} | \hat{\bm \pi} |\Psi_0^\text{PS}}, \label{eq:pi_ab_initio} \\
    \Braket{\hat{\bm r}}_{\Phi_0} &\equiv \Braket{\Phi_0 | \hat{\bm r} |\Phi_0}, \label{eq:r_ab_initio} \\
    \Braket{\hat{\bm \eta}^\text{nl}}_{\Phi_0} &\equiv \Braket{\Phi_0 | \left[ \hat{\bm r} , \sum_I \frac{\bm P_I}{M_I} \cdot \hat{\bm \Gamma}_I^\text{nl} \right] | \Phi_0}. \label{eq:nl_corr_ab_initio} 
\end{align}
Importantly, some care needs to be taken when using this ratio as a diagnostic -- see more details in Appendix B.

The many-body ground-state wavefunctions $\Ket{\Psi_0^\text{PS}}$ and $\Ket{\Phi_0}$ are evaluated within the independent-particle approximation using density functional theory at the generalized gradient approximation level, employing the Perdew-Burke-Ernzerhof exchange-correlation functional \cite{perdew_burke_ernzerhof_1996}. We compute the Nafie ratio twice for each diatomic molecule. In each calculation, one atom moves with velocity $\frac{P_I}{M_I}=0.001$ a.u. while the other remains stationary, with the roles of the two atoms interchanged between the two calculations. Note that the observables in Eqs.~\eqref{eq:pi_ab_initio}-\eqref{eq:nl_corr_ab_initio} for any vibrational or translational mode can be computed based on a linear combination of the observables obtained from these two calculations. 

The numerator of the Nafie ratio in Eq.~\eqref{eq:nafie_ratio_definition} is evaluated by diagonalizing the phase space Hamiltonian in Eq.~\eqref{eq:HPS_Gamma_nl}. In practice, we do not include the $\hat{\bm \Gamma}^{\prime \prime}_I$ terms above nor any $\hat{\bm \Gamma}_I \cdot \hat{\bm \Gamma}_I$ terms; our phase space electronic Hamiltonian is given by 
\begin{align}
    \hat{H}_\text{PS} = \hat{H}_\text{el} -i \hbar \sum_I \frac{\bm P_I \cdot \left( \hat{\bm \Gamma}^{\prime \text{local}}_I+\hat{\bm \Gamma}^{\prime \text{nl}}_I \right)}{M_I}. 
\end{align}
The space partitioning function $\theta_I \left( \bm r \right)$ in Eq.~\eqref{eq:Gamma_prime_local} is taken as the Hirshfeld weighting function \cite{hirshfeld1977, tkatchenko_scheffler_2009}
\begin{align}
    \theta_I \left( \bm r \right)=\frac{\rho_I\left( \bm r \right)}{\sum_J \rho_J\left( \bm r \right)},
\end{align}
where $\rho_I\left( \bm r \right)$ is the electron density for the isolated atom $I$.

For the denominator of the Nafie ratio in Eq.~\eqref{eq:nafie_ratio_definition} we perform BO calculations, where the first term is estimated with finite differences, using a time step of one atomic unit.

All calculations are performed with an in-house modified version of the \textsc{Quantum ESPRESSO} \cite{giannozzi_2009, giannozzi_2017} plane-wave code. We use optimized norm-conserving Vanderbilt pseudopotentials \cite{hammann_2013} obtained from the \textsc{pseudo-dojo} \cite{van_setten_rignanese_2018} online repository. The molecules are centered in a $38 \times 38 \times 38$ Bohr$^3$ cell in order to avoid spurious periodic image interactions, with a 100 Ry kinetic energy cutoff, and the Brillouin zone is sampled at a single point. We assume the experimental bond length \cite{johnson1999nist, nist_diatomic, minenkov2017ground, evans2000pure, brugh2002resonant}. Parts of the \textsc{Quantum ESPRESSO} implementation used in this work were developed with assistance from OpenAI Codex. All AI-assisted code was reviewed and validated by the authors.

\begin{figure}[htbp!]
    \centering
    \includegraphics[width=\linewidth]{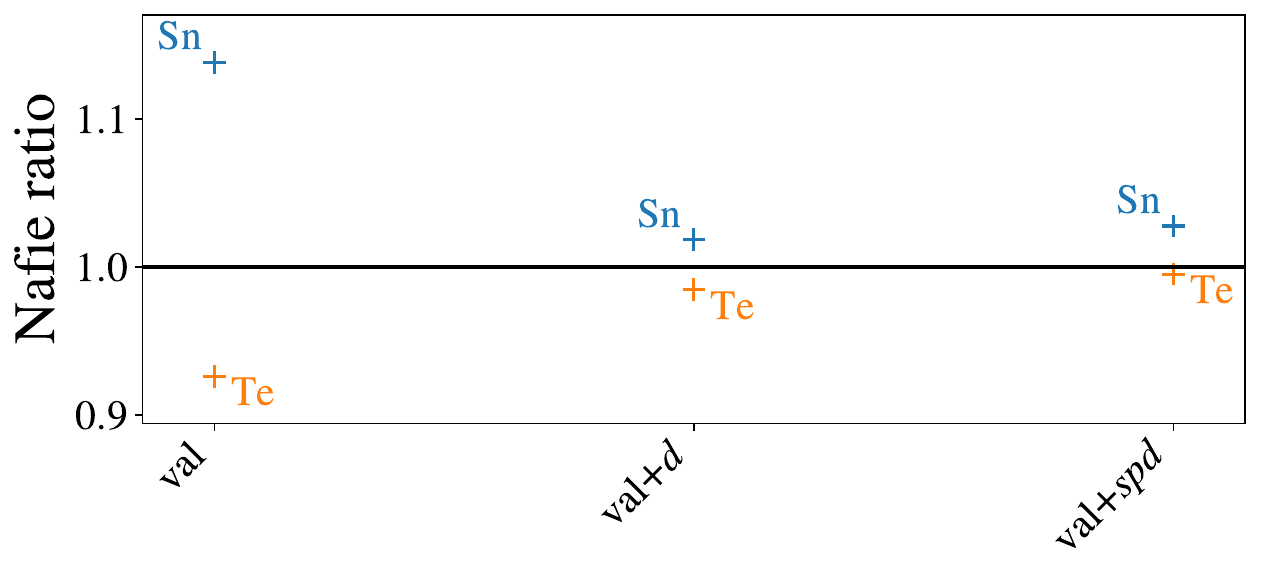}
    \caption{Same as Fig.~\ref{fig:nafie_ratio}, but for SnTe with different number of electrons treated explicitly in the pseudopotentials: val (outermost valence shell only), val+$d$ (outermost valence shell plus $d$ electrons of second outmost shell), and val+$spd$ (two outermost shells).}
    \label{fig:pp_nafie_ratio}
\end{figure}

\section{Results and discussion}
\label{sec:results}

The Nafie ratio in the vertical axis of Fig.~\ref{fig:nafie_ratio} is plotted for each single-atom vibrational mode for each diatomic molecule. Using the exact derivative coupling, the Nafie ratio is exactly 1, which is represented by the solid black line. Using an approximate $\hat{\bm \Gamma}_I$, an excellent agreement with the ideal ratio of 1 is readily observed for the entire set of 17 molecules. The mean absolute deviation from the ideal ratio is only 6\%, with a maximal deviation of 20\%. These results highlight the strength of our phase space approach in retrieving a meaningful electronic momentum, an ingredient that is completely missing in the BO picture.

Importantly, we find that the effect of the non-local contribution in Nafie's equality, namely $\Braket{\hat{\bm \eta}^\text{nl}}_{\Phi_0}$, is sizable in many cases. Excluding it, the mean absolute deviation from the ideal ratio increases to 28\%, with deviations of up to 147\%. The data is given in Appendix C.

Finally, Fig.~\ref{fig:pp_nafie_ratio} demonstrates that the Nafie ratio is not substantially affected by the number of electrons treated explicitly in the pseudopotential. For SnTe, across three different pseudopotential configurations, the smallest explicit valence space (outermost valence shell only) exhibits slightly larger deviations from the ideal ratio compared to the larger explicit valence spaces (which include electrons from the second outermost shell). Nevertheless, the agreement remains very good in all cases.

\section{Conclusions and Outlook}
\label{sec:conclusions}

We have introduced a framework for phase space electronic structure calculations with pseudopotentials, based on two new non-local ingredients, $\hat{\bm \Gamma}^{\prime \text{nl}}_I$ and $\hat{\bm \Gamma}^{\prime \prime \text{nl}}_I$, that guarantee conservation of kinetic linear and angular momentum respectively. This development represents an important step toward applying phase space theory within pseudopotential-based approaches, which underpin a large fraction of modern solid-state electronic structure calculations. It thereby opens the door to applications of phase space theory to non-adiabatic electron--phonon phenomena, including Einstein--de Haas physics and hopefully superconductivity.

\section*{Acknowledgments}

G.O. is grateful for support from the Council for Higher Education Fellowship for Postdoctoral Training in Quantum Science and Technology. This work was supported by the U.S. Air Force Office of Scientific Research (USAFOSR) under Grant No.~FA9550-23-1-0368. 

\section*{Data Availability}

The data that support the findings of this article are available from the authors upon reasonable request.

\section*{Appendix A: Proof of Eqs.~\NoCaseChange{\eqref{eq:Gamma_constr2_new}} and \NoCaseChange{\eqref{eq:Gamma_constr4_new}}}

To prove Eqs.~\eqref{eq:Gamma_constr2_new} and \eqref{eq:Gamma_constr4_new}, it is enough to show that the new ingredient introduced in $\hat{\bm \Gamma}_I^\text{nl}$, namely $\left[ \hat{v}_I^\text{nl}, \hat{\bm r} \right]$, is translationally invariant and transforms covariantly as a vector under global rotations. These facts follow immediately from the translational and rotational invariance of $\hat{v}_I^\text{nl}$, expressed as
\begin{align}
    \label{eq:joe1}
    &\left[ -i \hbar \sum_I \frac{\partial}{\partial \bm R_I} + \hat{\bm p},  \hat{v}_J^\text{nl} \right] = 0, \\
    \label{eq:joe2}
    &\left[ -i \hbar \sum_I \left( \bm R_I \times \frac{\partial}{\partial \bm R_I} \right)_\alpha + \hat{l}_\alpha + \hat{s}_\alpha, \hat{v}_J^\text{nl} \right] = 0.
\end{align}
Using Eqs. \eqref{eq:joe1}-\eqref{eq:joe2}, one can easily show that:
\begin{align}
    \left[ -i \hbar \sum_I \frac{\partial}{\partial \bm R_I} + \hat{\bm p}, \left[ \hat{v}_J^\text{nl}, \hat{\bm r} \right]  \right] = 0
\end{align}
and
\begin{align}
    \left[ -i \hbar \sum_I \left( \bm R_I \times \frac{\partial}{\partial \bm R_I} \right)_\alpha + \hat{l}_\alpha + \hat{s}_\alpha, \left[ \hat{v}_J^\text{nl}, \hat{r}_\beta \right]  \right] \nonumber \\
    = i \hbar \sum_\gamma \varepsilon_{\alpha \beta \gamma}  \left[\hat{v}_J^\text{nl}, \hat{r}_\gamma \right],
\end{align}
which follow from translational invariance and rotational covariance of $\hat{\bm r}$.

\section*{Appendix B: The Nafie ratio as a diagnostic}

\begin{figure}[htbp!]
    \centering
    \includegraphics[width=\linewidth]{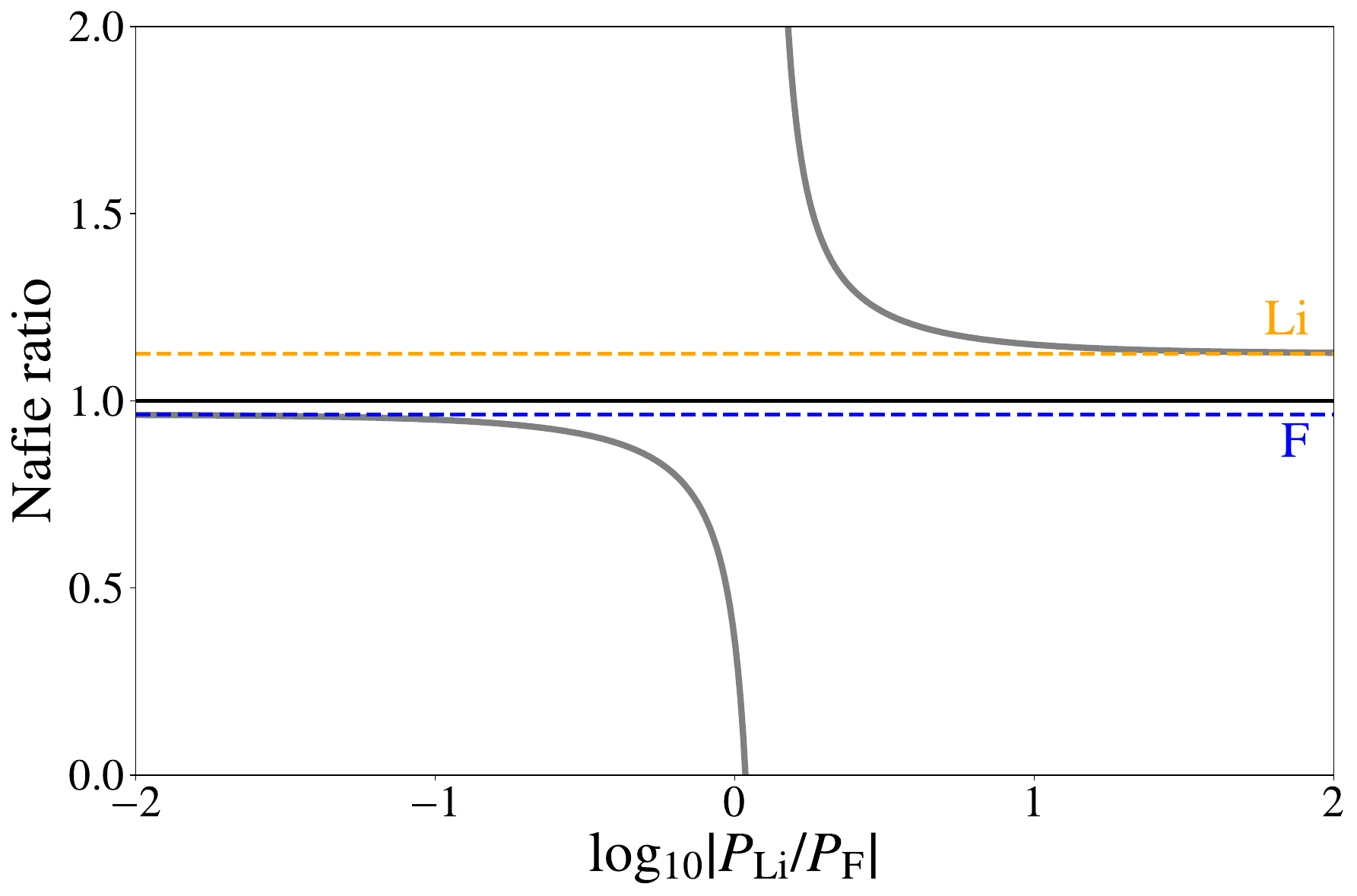}
    \caption{The Nafie ratio of a vibrating LiF molecule as a function of $\text{log}_{10} \left| P_\text{Li} / P_\text{F} \right|$. The solid black line corresponds to a ratio of unity, which is the ideal ratio in the limiting cases where only the Li (F) atom moves, represented by the dashed orange (blue) line.}
    \label{fig:LiF}
\end{figure}

The Nafie ratio in Eq.~\eqref{eq:nafie_ratio_definition} is not always a reliable diagnostic for validating Nafie's equality. Consider the case in which both the numerator and denominator of Eq.~\eqref{eq:nafie_ratio_definition} vanish exactly. Although Nafie's \emph{equality} is satisfied ($0=0$), the Nafie \emph{ratio} is undefined ($\frac{0}{0}$). Moreover, if there is any slight disagreement or error, one expects to find a divergence. Thus, for a set of nuclear momenta for which one expects a vanishing electronic momentum and no net change in the electronic dipole moment, the ratio should be unstable.

To illustrate this point, we consider the representative case of vibrations in LiF and vary the ratio between the nuclear momenta of the two atoms. In Fig.~\ref{fig:LiF}, we plot the Nafie ratio in Eq.~\eqref{eq:nafie_ratio_definition} against the quantity $x = \text{log}_{10} \left| P_\text{Li} / P_\text{F} \right|$. The two limiting cases in which one atom moves while the other remains fixed correspond to the limits $x \rightarrow \pm \infty$; these are the cases reported in the main text, for which a Nafie ratio of unity provides an appropriate diagnostic of agreement with Nafie's equality. As $x$ approaches $\sim0.1$, both the numerator and denominator of the Nafie ratio approach zero but not identically, rendering the ratio ill-defined and causing it to diverge. To avoid this behavior, in the main text, we have restricted our analysis to the limiting cases of single-atom translations, for which both the numerator and denominator remain sizable. We emphasize that the divergence in Fig.~\ref{fig:LiF} does not indicate a violation of Nafie's equality; rather, it reflects a limitation of the ratio as a diagnostic.

\section*{Appendix C: Effect of excluding $\Braket{\hat{\bm \eta}^\text{nl}}$ from the Nafie ratio}

\begin{figure*}[htbp!]
    \centering
    \includegraphics[width=\linewidth]{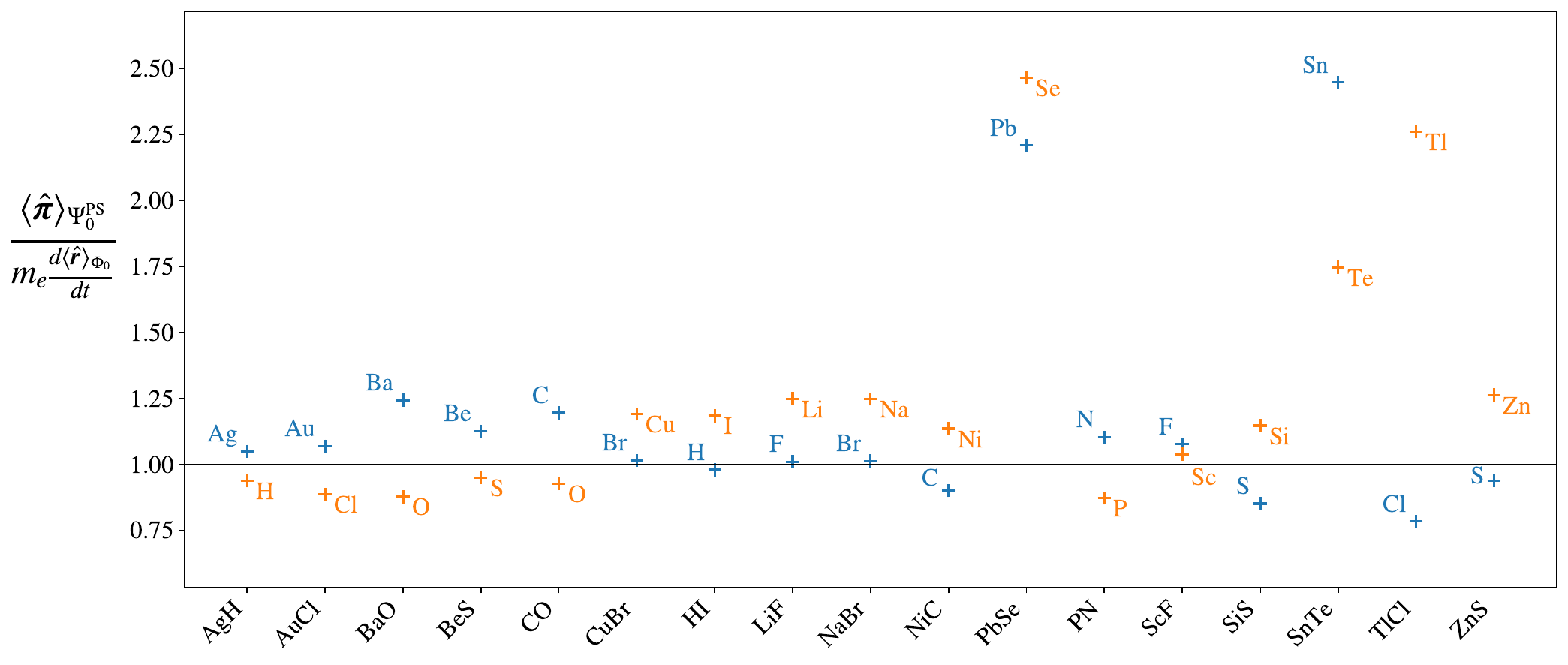}
    \caption{Same as Fig.~\ref{fig:nafie_ratio}, but the ratio on the vertical axis excludes $\Braket{\hat{\bm \eta}^\text{nl}}$.}
    \label{fig:nafie_partial_ratio}
\end{figure*}

Fig.~\ref{fig:nafie_partial_ratio} demonstrates the importance of the term associated with the non-local part of the pseudopotentials, $\Braket{\hat{\bm \eta}^\text{nl}}_{\Phi_0}$, for consistency with Nafie's equality. Excluding it, the deviations from the ideal ratio of unity increase substantially in several cases relative to the results shown in Fig.~\ref{fig:nafie_ratio}. A quantitative comparison between the two cases is provided in the main text.

\bibliography{references.bib}

\clearpage

\end{document}